\documentclass[12pt]{article}
\usepackage{latexsym}
\usepackage{amsmath}
\usepackage[dvips]{graphicx}
\usepackage{textcomp}

\begin{document}

\title{{\Large Reply to the "Comment on 'Piezonuclear decay of Thorium' [Phys. Lett. A
373 (2009) 1956]" [Phys. Lett. A 373 (2009) 3795] by G. Ericsson et
al.}}
\author{{\large Fabio Cardone}$^{1}${\large , Roberto Mignani}$^{2-4}$%
{\large \ and Andrea Petrucci}$^{5}$ \\
$^{1}$Istituto per lo Studio dei Materiali Nanostrutturati (ISMN -- CNR)\\
Via dei Taurini - 00185 Roma, Italy\\
$^{2}$GNFM, Istituto Nazionale di Alta Matematica "F.Severi"\\
\ Citt\`{a} Universitaria, P.le A.Moro 2 - 00185 Roma, Italy\\
$^{3}$Dipartimento di Fisica \textquotedblright E.Amaldi\textquotedblright ,
Universit\`{a} degli Studi \textquotedblright Roma Tre\textquotedblright \\
\ Via della Vasca Navale, 84 - 00146 Roma, Italy\\
$^{4}$I.N.F.N. - Sezione di Roma III\\
$^{5}$I.N.F.N. Laboratories - Frascati, Roma, Italy}
\maketitle
\date{}

\begin{abstract}
{\small In a paper appearing in this issue of Phys. Lett. A, Ericsson et al.
raise some critical comments on the experiment [Phys. Lett. A \ 373 (2009)
1956] we carried out by cavitating a solution of thorium-228, which
evidenced its anomalous decay behaviour, thus confirming the results
previously obtained by Urutskoev et al. by explosion of titanium foils in
solutions. In this letter, we reply to these comments. In our opinion, the
main shortcomings of the criticism by the Swedish authors are due to their
omitting of inserting our experiment in the wider research stream of
piezonuclear reactions, and to the statistical analysis they used, which
does not comply with the rules generally accepted for samples with small
numbers. However, apart from any possible theoretical speculation, there is
the basic fact that two different experiments (ours and that by Urutskoev et
al.), carried out independently and by different means, highlight an
analogous anomaly in the decay of thorium subjected to pressure waves. Such
a convergence of results shows that it is worth to further carry on
experimental investigations, in order to get either a confirmation or a
disproof of the induced-pressure anomalous behaviour of radioactive nuclides
even different from thorium. }
\end{abstract}

\newpage

\section{Introduction}

In an article appearing in the present issue of {\it Physics Letters A}$%
^{(1)}$, G.Ericsson et al. make some comments on our paper [2], in which we
report the results of an experiment investigating the possible effects of
cavitation on Thorium 228. It is just the aim of this letter to reply to
such comments. We are somehow grateful to the above Authors, because this
gives us the occasion to clarify some points of [2] which seemingly have
been misunderstood, and to deepen other ones.

In our opinion, the main shortcoming of paper [1] is the ignorance of the
bibliography concerning the subject of piezonuclear reactions (refs. [4-15]
of our article, we quote here again$^{(3-15)}$). This prevented Ericsson et
al. from putting our work [2] in this wider research stream, thus isolating
it from its natural context and misinterpreting its aims and results
(although we stated them in a quite clear way in the Introduction of [2]).

In short, piezonuclear reactions are nuclear reactions (of a new kind)
induced in liquids by ultrasound cavitation$^{(3-8)}$ and in solids by
mechanical pressing and consequent brittle failure$^{(9)}$. Such processes
give rise to transformations of elements$^{(3-5)}$ and to emission of
neutrons$^{(7,9,15)}$, {\em without significant emission of gamma radiation.}
Moreover, piezonuclear reactions are characterized by a marked threshold
behaviour in the supplied energy, and occur {\em in stable elements.}
Therefore, they do not belong to the research stream aimed at inducing {\em %
usual} nuclear reactions in unstable elements or deuterated compounds$%
^{(16-19)}$ by means of pressure waves (in particular ultrasounds and
cavitation), but constitute new phenomena involving the nucleus. A possible
theoretical interpretation of such new processes and of their features (in
terms of a spacetime deformation) has been given in refs.[6,8].

A research stream parallel to ours is given by the experiments, carried out
by Russian teams at Kurchatov Institute and at Dubna JINR, on the effects of
electric explosion of titanium foils in liquids$^{\left( 10-13\right) }$.
Analogous experiments are presently being performed at the Nantes GeM
laboratory$^{(14)}$. Besides observing transformation of elements (like in
our experiments [3-5]), the Russian researchers ascertained a violation of
the secular equilibrium of Thorium 234.

It was spontaneous for us to formulate the hypothesis that piezonuclear
reactions are at the basis of Russian results, namely that the explosion of
titanium foils gives rise to pressure waves in liquids able to generate
effects similar to cavitation. The simplest way to verify such a hypothesis
was just to carry out the experiment described in [2], i.e. to subject to
cavitation not stable nuclides (as we did in our previous experiments$%
^{(3-9,15)}$), but a radioactive one, we chose as $Th^{228}$ in order to
compare our results with the Russian ones.

This is the general context in which the experiment of paper [2] must be
inserted in order to fully appreciate its results. Our claim about the
piezonuclear decay of Thorium 228 is not based only on the experiment [2],
but is corroborated by our previous experiments and by those of the Russian
researchers. In particular, the statement in the Abstract of [1] that "..%
{\it a number of additional tests [that] should be made in order to improve
the quality of the study and test the hypothesis of so-called piezonuclear
reactions"}$^{(1)}${\it \ }is invalidated by the results of the experiments
[3-9,15], which show a quite clear evidence for reactions induced by
pressure (even neglecting the Russian experiments).

Therefore, the comments by Ericsson et al. apply at most to the results of
ref.[2]. Then, let us discuss in detail their main objections.

\section{Reply to Comments}

\subsection{Placement of detectors}

Obviously, the detectors CR39 where not placed {\em below} the cavitation
chamber, but {\em inside it on the bottom}. The error was due to an improper
use of the word "underneath" (which may in some case mean "inside on the
bottom"). During two years of discussions of the results with our colleagues
not directly involved in the experiment, it escaped our and our colleagues
revision of the manuscript, just because it was obvious for us and for them
that the CR39 were {\em inside} the vessel \ (by the way, their placement
was taken for granted also by the referee, who made no objections on this
point). Needless to say, we strongly apologize for this misprint, and are
grateful to Ericsson et al. for allowing us to clarify this fundamental
point.

\subsection{The $\protect\alpha $- traces on CR39}

With regard to the traces on the CR39 plates, we had already put in practice
the suggestion of the Authors of [1], namely "...{\it To investigate the
nature of the observed traces in the CR39 detectors, we suggest that the
authors conduct further background measurements, e.g., measurements during
cavitation of pure water, or with empty vessels, but in other respects
identical to what is done with the thorium solutions"}$^{(1)}${\it .} In
fact, in the experiments reported in refs.[6,15] we cavitated mere
bidistilled deionized water, by using two CR39, one inside and the other
below the cavitation chamber. We report here in Fig.1 the pictures of the
CR39 plates (already published in ref.[6], p.260).

\begin{figure}
\begin{center} \
\includegraphics[width=0.8\textwidth]{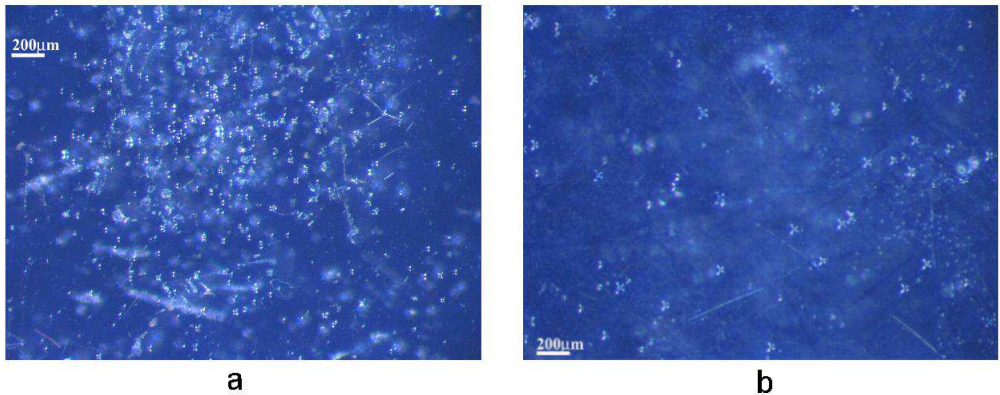} \caption{Showing the magnified central parts of the CR39 immersed in water,
on the bottom of the vessel (a) and of that below the vessel (b) for
cavitation of pure water. See refs.(6,15).}\label{traces_in_water}
\end{center}
\end{figure}

No trace on
them does vaguely resembles those observed in the thorium experiment$^{(2)}$%
. Moreover, the identification of the $\alpha $-traces was independently
made by dr. G. Cherubini, of the University of Roma "La Sapienza", formerly
at CAMEN (Center for Military Applications of Nuclear Energy), who prides
himself on decades-long experience in this field.

As to the number of detected $\alpha $-traces, our estimate of the
radioactivity of Thorium 228 is about one order of magnitude smaller than
that evaluated by the Authors of ref.[1]. But obviously this is not the
reason whereby we observed such a low number of decaying events. This is
instead due to the very objection raised by Ericsson et al. (see
Subsect.2.2) on the range of $\alpha $-particles. Clearly, since $\alpha $%
-particles in water have a very short range, the CR39 placed inside the
vessel is able to detect the decay of a nucleus of thorium only if it is
immediately near to (or even in touch with) the CR39 plate. We are amazed at
the fact that the Authors of [1] did not apply their own objection, based on
the range of $\alpha $-particles, to understand the low number of observed
traces.

\subsection{Increased Thorium radioactivity and $\protect\gamma $-emission}

Another misunderstanding of Ericsson et al. is to deduce that, since there
was a halving of the Thorium activity (and of the Thorium content in the
solutions) due to cavitation, this would imply an increase of thorium
radioactivity. Then, as a consequence, {\it "..the detectors monitoring the
cavitated solutions should not show fewer but four orders of magnitude more
events"}$^{(1)}$. This interpretation is based on the assumption that
thorium transformation proceeds through the usual decay channels, whereas in
our interpretation --- supported by our previous experiments --- thorium is
halved by means of piezonuclear reactions without increasing its
radioactivity.

By this token, the test suggested in [1] in order to check the presumed
increase of radioactivity of thorium, based on the analysis of $\gamma $
radiation, is invalidated by the fact that piezonuclear reactions occur
without $\gamma $ emission. We fully agree with Ericsson et al. that this
\textquotedblleft {\it \ldots would constitute a second extraordinary claim
which would need separate careful study, documentation and
verification.\textquotedblright }$^{(1)}$. The point is that we already
verified such an "extraordinary" fact in the previous fifteen years of
experiments$^{(3-9,15)}$ (let us also stress that such a feature, namely the
absence of $\gamma $ radiation is present also in the Russian experiments$%
^{(12,13)}$, and in some experiments of Low Energy Nuclear Reactions$^{(20)}$%
).

\subsection{Statistical analysis}

Concerning the statistical analysis, it was not our aim to carry out a
precise inferential statistical analysis in order to ascertain with a very
low level of significance if the reduction of radioactivity observed in the
cavitated samples were due not to \ chance, but to cavitation itself.
Actually, as already stressed, our paper [2] must be inserted in the context
of the evidences from other analogous experiments, in which cavitation was
shown to induce "anomalous" nuclear effects.

However, by going into details, it is not correct from a methodological
viewpoint --- as instead done in [1] --- to apply the test t of Student to
so scarce samples (in number of 3), although it holds for low samples but
anyway generally and basically not lower than 5$\div 7$ (see [21]).
Moreover, in paper [1], it is specified neither on what variable (counts or
concentrations) the test was performed, nor on what numbers and on what
calculation instrument it is based the value of 0.26 for p-value.

The Authors of [1] also state that the value of the p-value needed to
rejecting the null hypothesis must be lower than 0.01, whereas in general it
is usual to consider the value 0.05 as a limit\footnote{%
We recall that the "rule" of considering as significant p-values near to
zero is erroneously justified by misinterpreting the p-value as the
probability that the null hypothesis is true. For such a point and the
so-called Jeffreys-Lindley paradox, see ref.[21].}. However, even a so
stringent limit does not exist in any kind of inferential analysis, and the
assumed limiting values are largely conventional$^{(22)}$.

Therefore, we decided to carry out, in our turn, the t-test, deeming it the
most suited to the present case, although in presence of few data.

We utilized the software environment R, the most used by the international
scientific community for statistical computing, so anyone can easily test
our results.

By using the two-tailed t-test, with null hypothesis that both the cavitated
and the uncavitated sample belong to the same distribution and with
alternative hypothesis that the cavitated sample has values lower than the
uncavitated one, we found, by simply executing the t-test function, a
p-value of 0.062 on concentrations and of 0.065 on counts.

These values are definitely lower than those reported by Ericsson et al.. We
can therefore clearly state that the probability that the observed
differences, by accepting the null hypothesis, are due to chance, is only
6\%. Then, it is reasonable to reject the null hypothesis and to accept the
alternative hypothesis with a significance level of 6\%, very near to the
5\% usually accepted by the scientific community.

Let us notice that for scarce samples, as in the present case, it is even
more difficult, compared to bigger samples, rejecting the null hypothesis.
Therefore, the statistical result we obtained from the data of our
experiment has actually a great statistical meaning. This largely balances
also the (not low) inhomogeneity of the initial concentrations of all used
(both cavitated and reference) samples. It must be also taken into account
the difficulties of getting very uniform concentrations, owing to the
environmental and sanitary dangerousness of thorium. Then, although all the
laboratory expedients have been utilized in order to get the maximum
homogeneity in the concentrations of all samples, the obtained inhomogeneity
is actually an instrumental constraint inherent to the experiment itself.

In our opinion, apart from the initial inhomogeneity of the samples
(controlled and optimized compatibly with the experimental and technical
constraints), what does matter and was indeed observed is the halving of the
thorium content in the cavitated samples with respect to the uncavitated
ones, in agreement with the result obtained from the traces on the CR39
detectors.

\section{Conclusions}

At the light of the above reply to the comments by Ericsson et al. , we can
state that our experiment [2] has hit the two main aims we proposed to
ourselves, namely: 1) to verify the effect of pressure waves on thorium
decay, in agreement with the findings by Urutskoev et al.; 2) to confirm the
role of piezonuclear reactions in such a decay.

In our opinion, the main shortcomings of the criticism by the Swedish authors%
$^{(1)}$ are due to their omitting of inserting our experiment in the wider
research stream of piezonuclear reactions, and to the statistical analysis
they used, which does not comply with the rules generally accepted for
samples with very small numbers. However, apart from any possible
theoretical speculation, there is the basic fact that two different
experiments (ours and that by Urutskoev et al.), carried out independently
and by different means, highlighted an analogous anomaly in the decay of
thorium subjected to pressure waves. Such a convergence of results shows
that it is worth to further carry on experimental research, in order to get
either a confirmation or a disproof of the anomalous behaviour of
radioactive nuclides (even different from thorium) induced by pressure. On
this point, we fully agree with Ericsson et al. on the fact that our results
must be corroborated and confirmed by further and more refined
investigations. It was just one of the scopes of our paper [2] to stimulate
the scientific community in this sense, also on account of the possible
technology and application spin-off of our results. And, after all, an
experiment can only be really confirmed or confuted by another experiment.

{\it Acknowledgments.} We warmly thank W. Perconti, of the Data Analysis
Branch ISPRA (High Insttute for Enviromental Research), for very useful
discussions and enlightening comments..

\end{document}